\documentstyle[aps,preprint,amssymb,amscd,epsfig]{revtex}

\begin{document}

%%%%%%%%%%%%% Macro Definitions %%%%%%%%%%%%%%%%%%%%%%%%%%%%%%%%%%%%%%%%%%%%%%%

\def\tildeT{\tilde T}
\def\tildeC{\tilde C}
\def\tildeS{\tilde S}
\def\tildeP{\tilde P}

\def\F{F^{B\to T}}

\def\calA{{\cal A}}
\def\calB{{\cal B}}
\def\calO{{\cal O}}
\def\calX{{\cal X}}
\def\calY{{\cal Y}}
\def\calZ{{\cal Z}}

\def\Bbar{{\bar B}}

\def\vecp{{\vec p}}

%------------ PRD Convention --------------------------------------

\def\etal{{\it et al.}}
\def\ibid#1#2#3{{\it ibid.} {\bf #1}, #3 (#2)}
\def\epjc#1#2#3{Eur. Phys. J. C {\bf #1}, #3 (#2)}
\def\ijmpa#1#2#3{Int. J. Mod. Phys. A {\bf #1}, #3 (#2)}
\def\mpl#1#2#3{Mod. Phys. Lett. A {\bf #1}, #3 (#2)}
\def\npb#1#2#3{Nucl. Phys. {\bf B#1}, #3 (#2)}
\def\plb#1#2#3{Phys. Lett. B {\bf #1}, #3 (#2)}
\def\prd#1#2#3{Phys. Rev. D {\bf #1}, #3 (#2)}
\def\prl#1#2#3{Phys. Rev. Lett. {\bf #1}, #3 (#2)}
\def\rep#1#2#3{Phys. Rep. {\bf #1}, #3 (#2)}
\def\zpc#1#2#3{Z. Phys. {\bf #1}, #3 (#2)}
\def\ibid#1#2#3{{\it ibid}. {\bf #1}, #3 (#2)}

%%%%%%%%%%%%%% COVER PAGE %%%%%%%%%%%%%%%%%%%%%%%%%%%%%%%%%%%%%%%%%%%%%%%%%%%%%

\title{Nonleptonic two-body charmless $B$ decays involving a tensor meson in
ISGW2 model}
\author{C.~S. Kim\footnote{cskim@mail.yonsei.ac.kr},
Jong-Phil Lee\footnote{jplee@phya.yonsei.ac.kr},
Sechul Oh\footnote{scoh@post.kek.jp}}
\address{Department of Physics and IPAP, Yonsei University, Seoul, 120-749, Korea}

%\tighten
\maketitle

\begin{abstract}

Nonleptonic charmless $B$ decays into a pseudoscalar $(P)$ or a vector $(V)$
meson accompanying a tensor $(T)$ meson are re-analyzed.
We scrutinize the hadronic uncertainties and ambiguities of the form factors
which appear in the literature.
The Isgur-Scora-Grinstein-Wise updated model (ISGW2) is adopted to
evaluate the relevant hadronic matrix elements.
We calculate the branching ratios and CP asymmetries for various $B\to P(V)T$
decay processes.
With the ISGW2 model, the branching ratios are enhanced by about an order
of magnitude compared to the previous estimates.

\end{abstract}
\pagebreak

%%%%%%%%%%%%%%%%%%%%%%%%%%%%%%%%%%%%%%%%%%%%%%%%%%%%%%%%%%%%%%%%%%%%%%%%%%%%%%%
\section{Introduction}
%%%%%%%%%%%%%%%%%%%%%%%%%%%%%%%%%%%%%%%%%%%%%%%%%%%%%%%%%%%%%%%%%%%%%%%%%%%%%%%

With the beginning of the $B$-factory epoch, a tremendous amount of
experimental data on $B$ decays start to provide new bounds on previously known
observables with an unprecedented precision as well as an opportunity to see
very rare decay modes for the first time.
Experimentally several tensor mesons have been observed \cite{PDG},
such as the isovector $a_2$(1320), the isoscalars $f_2$(1270),
$f_2^{\prime}$(1525), $f_2$(2010), $f_2$(2300), $f_2$(2340),
$\chi_{c2}(1P)$, $\chi_{b2}(1P)$ and $\chi_{c2}(2P)$, and the
isospinors $K_2^*$(1430) and $D_2^*$(2460).
The measured
branching ratios for $B$ decays involving a pseudoscalar $(P)$ or a
vector ($V$), and a tensor meson ($T$) in the final state provide only upper
bounds.  For example \cite{PDG},
\begin{eqnarray}
{\cal B} (B^{+(0)} \rightarrow D_2^*(2460)^{0(-)} \pi^+) &<& 1.3
(2.2) \times 10^{-3},   \nonumber
\\ {\cal B} (B^{+(0)} \rightarrow D_2^*(2460)^{0(-)} \rho^+) &<& 4.7
(4.9) \times 10^{-3},   \nonumber
\\ {\cal B} (B^+ \rightarrow K_2^*(1430)^0 \pi^+) &<& 6.8 \times
10^{-4},   \nonumber
\\ {\cal B} (B^{+(0)} \rightarrow K_2^*(1430)^{+(0)} \rho^0) &<& 1.5
(1.1) \times 10^{-3},   \nonumber
\\ {\cal B} (B^{+(0)} \rightarrow K_2^*(1430)^{+(0)} \phi) &<& 3.4
(1.4) \times 10^{-3},   \nonumber
\\ {\cal B} (B^+ \rightarrow \pi^+ f_2(1270)) &<& 2.4 \times
10^{-4},   \nonumber
\\ {\cal B} (B^+ \rightarrow \rho^0 a_2(1320)^+) &<& 7.2 \times
10^{-4},   \nonumber
\\ {\cal B} (B^0 \rightarrow \pi^{\pm} a_2(1320)^{\mp}) &<& 3.0 \times
10^{-4}.
\end{eqnarray}
In particular, the process $B \to K_2^* \gamma$ has been observed
for the first time by the CLEO Collaboration with a branching
ratio of $(1.66^{+0.59}_{-0.53} \pm 0.13) \times 10^{-5}$
\cite{CLEO}, and by the Belle Collaboration with
$\calB(B\to K_2^*\gamma)=(1.26\pm0.66\pm0.10)\times 10^{-5}$ \cite{Belle0}.
\par
{}From the theoretical point of view, nonleptonic two-body decays are quite difficult
to deal with because of our poor understandings of the nonperturbative nature
of the hadronic matrix elements.
The factorization hypothesis is a widely used scheme \cite{Neubert}.
Within the framework of generalized factorization, it can be easily shown that
$\langle 0|j^\mu|T\rangle=0$, where $j^\mu$ is the $V-A$ current \cite{Oh1,Oh2}.
Since there are no terms proportional to the tensor decay constant $f_T$ at
the amplitude level, the decay amplitudes for $B\to P(V)T$ can be considerably
simplified compared to those for other two-body $B$ decays such as
$B\to PP,~PV,~{\rm or}~VV$.
This is a great advantage for our theoretical predictions.
Given the factorization assumption, therefore, the hadronic uncertainties are
condensed to the $B\to T$ form factors.

Based on the nonrelativistic quark model, Isgur, Scora, Grinstein and Wise
(ISGW) suggested the weak transition form factors of $B\to X(=q{\bar d})$,
where $X$ is in $1(2) ^{1,3}S_0$, $1 ^3P_{2,1,0}$, $1 ^1P_1$ states,
in analyzing semileptonic $B\to X\ell{\bar\nu}$ decays \cite{ISGW}.
There exist a few works on two-body hadronic $B$ decays
\cite{Katoch,Castro,Munoz} that involve a tensor meson $T$ ($J^P = 2^+$) in
the final state using the ISGW model, together with the factorization ansatz;
those works considered only the tree diagram contributions.

In recent works \cite{Oh1,Oh2}, we have analyzed charmless $B\to P(V)T$ decays
in the framework of both flavor SU(3) symmetry and the generalized
factorization.
The works included all the penguin operators in the effective Hamiltonian,
and adopted the ISGW model.
The main idea of the ISGW model is that the weak transition matrix elements
are calculable in the nonrelativistic limit where the constituents are
very heavy compared to the typical QCD scale $\Lambda_{\rm QCD}$.
One crucial hypothesis of the model is that a smooth extrapolation to the real
physical regime is acceptable.
With its Gaussian predictions, however, the ISGW model
underestimates the form factors
at high $q^2$ in the semileptonic decays \cite{ISGW2}.
In the two-body hadronic decays the form factors are severely suppressed,
yielding the typical branching ratios of $B\to P(V)T$ to be
$\calO(10^{-8})\sim\calO(10^{-7})$ \cite{Oh1,Oh2}.
The Belle Collaboration is currently searching for some $B\to PT$ modes and
their (very) preliminary result indicates that the branching ratios for those
modes may not be very small compared to  $B\to PP$ modes \cite{Belle}.

The authors of \cite{Katoch} first pointed out the above mentioned problem
in the calculation of the branching ratios of $B\to P(V)T$.
They suspected the reliability of the exponentially decreasing behavior of
the form factors, and used the form factors calculated at maximum momentum
transfer $t_m\equiv (m_B-m_T)^2$.
But the shift of the kinematical point from $m_{P(V)}^2$ to $t_m$ yields a
tremendous amount of enhancement in the branching ratios.
As we shall see in Sec.\ II, the branching ratios can increase by two orders of
magnitude.
Though the use of $t_m$ instead of $m_{P(V)}^2$ may fit the data
phenomenologically, its origin is also less justified.

The advent of the heavy quark effective theory (HQET) based on a deeper
understanding of the heavy quark
symmetry (HQS) allowed to consolidate the foundations of a new model, upgrading
ISGW into its second version (ISGW2) \cite{ISGW2}.
In this work, we update our previous analyses with the ISGW2 model for the charmless
$B\to P(V)T$ decay processes \cite{new-charmed}.
New features and merits of ISGW2 are the byproducts of the HQS.
For instance, the relations between form factors are respected and the weak
currents in the effective theory are matched to those in full QCD via
the renormalization group flow \cite{HQET}.
Though the HQS cannot replace the role of model calculations, it also alleviates
some ambiguities in the original ISGW model,
such as the absence of the relativistic corrections.

The ISGW2 modifies the form factors in a more realistic manner.
Most importantly, the Gaussian factor is changed into a polynomial.
The sensitivity to the kinematical point of the form factors now becomes rather
moderate, and their high-$q^2$ behavior fits the data well \cite{ISGW2}.
%In ISGW2, the form factors increase as a whole.
The values of the form factors are not so suppressed as in ISGW.
The increase in the branching ratios is significant, and the enhanced
branching ratios are of order ${\cal O}(10^{-6})$, Belle and BaBar can now check.
On the other hand, the CP asymmetry and the ratios of the branching ratios
of $B\to VT$ and $B\to PT$ would not be affected so much.
They reduce model dependence and will check the general framework of the
factorization and flavor SU(3) symmetry.
\par
The paper is organized as follows.
Section II parameterizes the hadronic matrix elements in the framework of
generalized factorization.
The ISGW2 model is adopted to evaluate the form factors.
In Sec.\ III, the branching ratios, the CP asymmetries, and the $VT/PT$ ratio
$\calB(B\to VT)/\calB(B\to PT)$ are calculated by using ISGW2 form factors.
The meanings of our numerical results are also discussed.
We conclude the analysis in Sec.\ IV.

%%%%%%%%%%%%%%%%%%%%%%%%%%%%%%%%%%%%%%%%%%%%%%%%%%%%%%%%%%%%%%%%%%%%%%%%%%%%%%%
\section{Form Factors and ISGW2}
%%%%%%%%%%%%%%%%%%%%%%%%%%%%%%%%%%%%%%%%%%%%%%%%%%%%%%%%%%%%%%%%%%%%%%%%%%%%%%%

We refer to previous works \cite{Oh1,Oh2,ddo,Gronau,Gronau2,Dighe} for relevant
conventions and notations.
The main difficulty in theoretical predictions comes from the hadronic matrix
elements.
We adopt the factorization assumptions in $B\to P(V)T$, and then use the
ISGW2 model to evaluate $B\to T$ transition matrix elements.
In the factorization framework, the matrix elements for $B\to P(V)T$ are
parameterized as \cite{ISGW}
\begin{eqnarray}
\langle 0 | A^{\mu} | P \rangle &=& i f_P p_P^{\mu} ~,  \\
\langle 0 | V^{\mu} | V \rangle &=&  m_V f_V \epsilon^{\mu} ~,  \\
\langle T | j^{\mu} | B \rangle &=& i h(m_{P(V)}^2) \epsilon^{\mu \nu
\rho \sigma} \epsilon^*_{\nu \alpha} p_B^{\alpha} (p_B
+p_T)_{\rho} (p_B -p_T)_{\sigma} + k(m_{P(V)}^2) \epsilon^{* \mu \nu}
(p_B)_{\nu}  \nonumber \\
&\mbox{}&  + \epsilon^*_{\alpha \beta} p_B^{\alpha} p_B^{\beta} [
b_+(m_{P(V)}^2) (p_B +p_T)^{\mu} +b_-(m_{P(V)}^2) (p_B -p_T)^{\mu} ]~,
\label{formfactor}
\end{eqnarray}
where $j^{\mu} = V^{\mu} -A^{\mu}$.  $V^{\mu}$ and $A^{\mu}$
denote a vector and an axial-vector current, respectively.  $f_{P(V)}$
denotes the decay constant of the relevant pseudoscalar (vector) meson.
$p_B$ and $p_T$ denote the
momentum of the $B$ meson and the tensor meson, respectively.
Here $\epsilon^\mu (\epsilon^{\mu\nu})$ is the polarization vector (tensor)
of the vector (tensor) meson.
The polarization tensor $\epsilon^{\mu \nu}$ satisfies the following properties
\cite{epsilon}:
\begin{eqnarray}
&\mbox{}& \epsilon^{\mu \nu} (p_{_T}, \lambda) = \epsilon^{\nu
\mu} (p_{_T}, \lambda),
\\ &\mbox{}& p_\mu \epsilon^{\mu \nu} (p_{_T}, \lambda) = p_\nu \epsilon^{\mu \nu} (p_{_T},
\lambda) = 0,
\\ &\mbox{}& \epsilon^\mu_{~\mu} (p_{_T}, \lambda) = 0~,
\end{eqnarray}
where $\lambda$ is the helicity index of the tensor meson.
Note that, as argued in \cite{Oh1}, there are no amplitudes proportional
to $f_T\times ({\rm form~factor~for~}B\to P(V))$ since
\begin{equation}
\langle 0|j^\mu|T\rangle=p_\nu\epsilon^{\mu\nu}(p_T,\lambda)
 +p_T^\mu\epsilon^\nu_\nu(p_T,\lambda)=0~.
\end{equation}

The coefficients $h$, $k$, $b_\pm$ contain nonperturbative nature of the
$B\to T$ transition.
They are in general functions of the momentum transfer $t\equiv(p_B-p_T)^2$,
and are combined to express the form factors for $B\to T$, $\F(t)$.
Explicitly \cite{Oh1,Oh2,Oh},
\begin{equation}
\calA(B\to PT)\sim \F(m_P^2)~,~~~~~
\calA(B\to VT)\sim \epsilon^{*\alpha\beta}\F_{\alpha\beta}(m_V^2)~,
\end{equation}
where
\begin{mathletters}
\begin{eqnarray}
\F(m_P^2)&=&k(m_P^2)+(m_B^2-m_T^2)b_+(m_P^2)+m_P^2b_-(m_P^2)~,\\
\F_{\alpha\beta}(m_V^2)&=&\epsilon^*_\mu(p_B+p_T)_\rho
 \Big[ih(m_V^2)\cdot \epsilon^{\mu\nu\rho\sigma}
  g_{\alpha\nu}(p_V)_\beta(p_V)_\sigma
 +k(m_V^2)\cdot\delta^\mu_\alpha\delta^\rho_\beta\nonumber\\
&& +b_+(m_V^2)_\cdot(p_V)_\alpha(p_V)_\beta g^{\mu\rho}\Big]~.
\end{eqnarray}
\end{mathletters}
The ISGW(2) is designed to evaluate $h$, $k$, and $b_{\pm}$, based on the
nonrelativistic quark potential.
Originally, the ISGW model was introduced to see that the free-quark decay model
for $B\to X\ell{\bar\nu}$ might be deficient in the end point region where
the lepton energy is near its maximum.

One important demerit of the ISGW is that it underestimates the form factors
at high $q^2$ \cite{ISGW2}.
This deficiency is due to the Gaussian wave function, which is a direct
consequence of the classical potential.
Besides, the exponential factor makes it
unnatural that the decay rates are very sensitive to the kinematical points
where the two-body decay rate is determined.
Even a small uncertainty can result in a big fluctuation of the form factors.
As an illustration,
we list in Table \ref{FISGW} the values of the form factors calculated at
$q^2=m_\pi^2,~m_K^2$, and $t_m$ in the ISGW model (corresponding values in
ISGW2 are also given as comparisons) for $B\to PT$.
At $q^2=m_\pi^2$ or $m_K^2$, $\F\approx -0.03$ for all $T$.
However, at maximum momentum transfer, the values of the form factors increase
by almost an order of magnitude, e.g., $\F\approx -0.2$ for $T=a_2,~f_2,~f_2'$.
Consequently, the prediction of the branching ratios for $B\to PT$ increase by
about two orders of magnitude.
Typically, the branching ratios increase from $\calO(10^{-8})\sim\calO(10^{-7})$
to $\calO(10^{-6})\sim\calO(10^{-5})$.
For example, $\calB(B^+\to\pi^+ a_2^0)\simeq (35-45)\times 10^{-8}$ with
$\F(m_\pi^2)$, while the branching ratio amounts to
$\simeq (15-19)\times 10^{-6}$ when $\F(t_m)$ is used.
\par
In the ISGW2 model, all the advantages of the heavy quark symmetry (HQS) are
included.
Though the ISGW2 still needs model calculations,
the use of HQS considerably reduces the ambiguities which appeared in the original
ISGW when obtaining physical form factors.
The ISGW2 model changes the exponential factor of $\F$ into a polynomial
\cite{ISGW2},
\begin{eqnarray}
\label{1to2}
h,~k,~b_\pm&\sim&\exp[-({\rm const.})\times(t_m-t)]~~~~~({\rm ISGW})\nonumber\\
&\Rightarrow&
[1+({\rm const.})\times(t_m-t)]^{-N}~~~({\rm ISGW2})~,
\end{eqnarray}
where $N=3$ is a model parameter.
In addition, some relativistic corrections have clear advantages in ISGW2.
For example, the hyperfine splittings between $\Bbar-\Bbar^*$, $D-D^*$, which
were not taken into account in the original ISGW, have a natural origin of chromomagnetic
operators in HQET at order $1/m_Q$.
As will be seen in the next section,
all of the changes in ISGW2 are combined to increase the form factors.

On the other hand, the CP asymmetry
\begin{equation}
{\cal A}_{CP}=\frac{{\cal B}(B\to f)-{\cal B}({\bar B}\to{\bar f})}
              {{\cal B}(B\to f)+{\cal B}({\bar B}\to{\bar f})}~,
\end{equation}
where ${\cal B}(B\to f)$ is the branching ratio for a $B$ meson decaying into
a generic final state $f$,
and the $VT/PT$ ratio
\begin{eqnarray}
R_{V/P}&\equiv&
\frac{\calB(B\to VT)}{\calB(B\to PT)}\nonumber\\
&=&
[{\rm CKM~and~QCD~factors}]\times
\Bigg(\frac{f_V}{f_P}\Bigg)^2
\Bigg(\frac{|\vecp_V|}{|\vecp_P|}\Bigg)^5
\frac{\Big[\calX|\vecp_V|^2+\calY+\calZ/|\vecp_V|^2\Big]_{t=m_V^2}}
{2[m_B\F(m_P^2)]^2}~,
\label{vtoverpt}
\end{eqnarray}
where
\begin{mathletters}
\begin{eqnarray}
\calX&=&8m_B^4b_+^2~,\\
\calY&=&2m_B^2\big[6m_V^2m_T^2h^2+2(m_B^2-m_T^2-m_V^2)kb_++k^2\big]~,\\
\calZ&=&5m_T^2m_V^2k^2~,
\end{eqnarray}
\end{mathletters}
are both expected to be less model-dependent.
The reason is that the overall factor of exponential or polynomial of the
form factors in (\ref{1to2}) almost cancels out.
Considering the three degrees of freedom of vector mesons, it is quite natural
to guess $R_{V/P}\sim 3$ naively, just as in other analogous decay modes.
Measurements of $R_{V/P}$ in future experiments, therefore, will not only check
the model predictions but also examine the validity of the factorization scheme.
Since we include all the penguin operators in the effective Hamiltonian, it is
possible that the [CKM and QCD factors] $\ne 1$.

%%%%%%%%%%%%%%%%%%%%%%%%%%%%%%%%%%%%%%%%%%%%%%%%%%%%%%%%%%%%%%%%%%%%%%%%%%%%%%%
\section{Results and discussions}
%%%%%%%%%%%%%%%%%%%%%%%%%%%%%%%%%%%%%%%%%%%%%%%%%%%%%%%%%%%%%%%%%%%%%%%%%%%%%%%

In this section, we give the branching ratios and CP asymmetries for
$B\to P(V)T$ by using the ISGW2 model.
The results are summarized in Tables \ref{PS0B}-\ref{PS1A} for $B\to PT$ and
Tables \ref{VS0B}-\ref{VS1A} for $B\to VT$.
All the input parameters, such as decay constants, quark masses,
$\xi \equiv 1/ N_c$
($N_c$ denotes the effective number of color), and
CKM elements, etc. are the same as the previous ones of \cite{new-charmed}.
Note that the improved Wilson coefficients (WCs) for CKM angle 
$\gamma=65^\circ$ and $m_s(m_b)=100$ MeV are used \cite{Ali}.
The running quark masses (in MeV) at $m_b$ scale are used \cite{Dutta}:
$m_u=3.6$, $m_d=6.6$.
We give the results for different values of the parameter $\xi=0.1,~0.3,~0.5$.
In the framework of the QCD factorization and the perturbative QCD, $\xi$ can
be different for tree- and penguin-dominated processes.
But in this work where the generalized factorization is adopted, the universal
$\xi$ is assumed.
\par
Compared to the original ISGW results \cite{Oh1,Oh2}, the branching ratios
are enhanced by about one order of magnitude.
The enhancement in the branching ratio is not solely due to the change of
(\ref{1to2}).
Typically,
$$
\frac{[1+({\rm const.})\times(t_m-t)]^{-N}}
{\exp[-({\rm const.})\times(t_m-t)]}~ \gtrapprox ~ 3.
$$
In the $\Delta S=0$ case, the decay modes
$B^+\to\pi^+(\rho^+)a_2^0$, $B^+\to\pi^+(\rho^+)f_2$, and
$B^0\to\pi^+(\rho^+)a_2^-$ have relatively large branching ratios,
$\calO(10^{-6})\sim\calO(10^{-5})$.
On the other hand, $\calB(B^+\to\pi^0(\rho^0)a_2^+)$ is much smaller than
$\calB(B^+\to\pi^+(\rho^+)a_2^0)$ by several ten times, depending on the
input parameters.
The reason of the suppression is, as argued in \cite{Oh1}, that in the
factorization scheme the dominant contribution to the former arises from the
color-suppressed tree diagram ($C_T$), while the dominant one to the latter
arises from the color-favored tree diagram ($T_T$).
In $|\Delta S|=1$ modes, $B^{+(0)}\to\eta' K_2^{*+(0)}$ in $B\to PT$ and
$B^+\to K^{*+}a_2^0,~K^{*+}f_2,~K^{*0}a_2^+$,
$B^0\to K^{*+}a_2^-,~K^{*0}a_2^0,~K^{*0}f_2$ in $B\to VT$
have relatively larger branching ratios of $\calO(10^{-6})$.
\par
On the other hand, the branching ratios of $B^+\to\pi^0(\rho^0)a_2^+$,
$B^0\to\pi^0(\rho^0)a_2^0$, $B^0\to\pi^0(\rho^0)f_2$,
$B^0\to\pi^0(\rho^0)f_2'$ for $\xi=0.3$ are quite small compared to those
values for $\xi=0.1$ or $0.5$.
The tree diagrams of the processes are proportional to 
$a_2\equiv c_2+\xi c_1$, where $c_{1,2}$ are the effective WCs.
As argued in \cite{new-charmed}, the value of $a_2$ for $\xi=0.3$ is about
an order of magnitude smaller than that for $\xi=0.1$ or $0.5$, leading to
small branching ratios.
\par
Other properties of the branching ratios discussed in \cite{Oh1,Oh2} still hold
in ISGW2.
For example,
\begin{mathletters}
\begin{eqnarray}
2\calB(B^+\to\pi^+(\rho^+)~a_2^0)&\approx&\calB(B^0\to\pi^+(\rho^+)~a_2^-)~,\\
\calB(B^+\to\pi^0 a_2^+)&\ll&\calB(B^+\to\pi^+ a_2^0)~,\\
\calB(B^+\to\pi^0 K_2^{*+})&\approx&\calB(B^0\to\pi^0 K_2^{*0})~,\\
\calB(B^+\to\rho^0 a_2^{+(0)})&\approx&\calB(B^+\to\omega a_2^{+(0)}),\\
\calB(B^+\to\rho^0f_2^{(\prime)})&\approx&\calB(B^+\to\omega f_2^{(\prime)})~.
\end{eqnarray}
\end{mathletters}

Recently, it is reported that the branching ratio of
$\calB(B^+\to K^+ \pi^+\pi^-)=(55.6\pm5.8\pm7.7)\times 10^{-6}$ was
experimentally measured for the first time by Belle \cite{Belle}.
Two possible states for a $\pi^+\pi^-$ invariant mass around 1300 MeV are
suggested; $f_0(1370)$ and $f_2(1270)$.
Referring to our previous predictions \cite{Oh1} with the original ISGW model,
they concluded that the measurements would
provide evidence for a significant nonfactorizable contribution, if the peak
were due to $f_2(1270)$.
Now the branching ratio is enhanced as
$\calB(B^+\to K^+ f_2)\approx 10^{-8}~({\rm ISGW})
\Rightarrow 10^{-7}~({\rm ISGW2})$.
Though the enhancement dose not tell where the peak comes from, we can say
that nonfactorizable effects are considerably reduced if $\pi^+\pi^-$ is from
the resonance $f_2(1270)$.
\par
While the branching ratios become larger by about ten times in ISGW2 model,
the CP asymmetry $\calA_{CP}$ and the ratio $R_{V/P}$  remain almost unchanged.
The reason is that the model dependence nearly drops out, though not exactly,
in the ratios.
\par
As for the CP asymmetry, the modes
$B^+\to\eta^{(\prime)} a_2^+$, $B^0\to\eta a_2^0(f_2)$,
$B^{+(0)}\to\eta K_2^{*+(0)}$ in $B\to PT$ and
$B^+\to\rho^0(\omega)a_2^+$, $B^+\to K^{*+}a_2^0(f_2)$,
$B^0\to K^{*+}a_2^-$ in $B\to VT$ have relatively large $\calA_{CP}$'s and
$\calB$'s, just as in the ISGW case.
The modes of $B^+\to K^{(*)0}a_2^+$, $B^0\to K^{(*)0}a_2^0$,
$B^0\to K^{(*)0}f_2$, $B^0\to K^{(*)0}f_2'$ show vanishing CP asymmetry
because they have no tree-penguin interferences.

Next we consider the ratio $R_{V/P}$.
Some of the ratios are given in Table \ref{RVP}.
As expected in Sec.\ II, $R_{\rho/\pi}\sim 3$ irrespective of $m_s$ and
$\gamma$.
This is quite a reasonable tendency because of 3-helicity
states of vector mesons.

In cases where the strangeness changes ($|\Delta S|=1$), the ratio $R_{V/P}$
can be about 6, much larger than that of $\Delta S=0$ processes.
The reason is that the term [CKM and QCD factors] in Eq. (\ref{vtoverpt}) does not
disappear (when Fierzing the four-quark penguin operators, terms containing
(pseudo)scalar current survive only in $\calB(B\to PT)$), and the factors are
very different in $|\Delta S|=1$ and $\Delta S=0$ modes.
For $|\Delta S|=1$ decays, the term [CKM and QCD factors] amounts
to $\approx 2.5 \sim 4.5$, while it is $\approx 0.5\sim0.6$ in
$|\Delta S|=0$ decay modes.

%%%%%%%%%%%%%%%%%%%%%%%%%%%%%%%%%%%%%%%%%%%%%%%%%%%%%%%%%%%%%%%%%%%%%%%%%%%%%%%
\section{Conclusions}
%%%%%%%%%%%%%%%%%%%%%%%%%%%%%%%%%%%%%%%%%%%%%%%%%%%%%%%%%%%%%%%%%%%%%%%%%%%%%%%

Stimulated by the Belle measurements of possibly large branching ratios,
we have re-examined our previous works on the exclusive charmless decays
$B\to P(V)T$ in the frameworks of generalized factorization.
The main source of theoretical uncertainties comes from the hadronic matrix
elements.
Under the factorization hypothesis, the decay amplitudes proportional to the
tensor decay constant are absent, so the evaluation of the $B\to T$ transition
matrix elements is crucial.
The original ISGW model tends to underestimate the relevant form factors of
$B\to T$ and has some ambiguities in its model parameters.
The HQS-based ISGW2 model provides more reliable form factors.
Compared to the results of its previous version, ISGW2 predicts about 10 times
larger branching ratios.
The modes of $B^+\to\pi^+(\rho^+)a_2^0,~\pi^+(\rho^+)f_2$,
$B^0\to\pi^+(\rho^+)a_2^-$, and
$B^+\to K^{*+}a_2^0,~K^{*+}f_2,~K^{*0}a_2^+$,
$B^0\to K^{*+}a_2^-,~K^{*0}a_2^0,~K^{*0}f_2$
show conspicuously large branching ratios of order $\calO(10^{-6})$.

On the other hand, the CP asymmetry $\calA_{CP}$ and the fraction of
$R_{V/P}=\calB(VT)/\calB(PT)$ remain almost unchanged.
This is because the model dependence nearly cancels in those ratios.
\par
There still needs more embellishment to control the hadronic uncertainties.
Estimating the nonfactorizable effects, for example, will improve the
predictivity.
On the experimental side, the exclusive charmless decays
$B\to P(V)T$ can be carried out in details at hadronic $B$ experiments
such as BTeV and LHC-B, where more than $10^{12}$ $B$ mesons will be produced
per year as well as at present leptonic
asymmetric $B$ factories of Belle and BaBar.

%%%%%%%%%%%%%%%%%%%%%%%%%%%%%%%%%%%%%%%%%%%%%%%%%%%%%%%%%%%%%%%%%%%%%%%%%%%%%%
\begin{center}
{\large\bf Acknowledgements}\\[10mm]
\end{center}
%%%%%%%%%%%%%%%%%%%%%%%%%%%%%%%%%%%%%%%%%%%%%%%%%%%%%%%%%%%%%%%%%%%%%%%%%%%%%%
\noindent We thank F. Kr\"uger for careful reading of the manuscript and his
valuable comments.
The work of C.S.K. was supported in part by  CHEP-SRC
Program, Grant No. 20015-111-02-2 and Grant No.  
R02-2002-000-00168-0 from BRP  of
the KOSEF, and in part by  Grant No. 2001-042-D00022
of the KRF. The work of J.-P.L. was supported by the BK21 Program.
The work of S.O. was supported by the KRF
Grants, Project No. 2001-042-D00022.

%%%%%%%%%%%%%%%%%%%%%%%%%%%%%%%%%%%%%%%%%%%%%%%%%%%%%%%%%%%%%%%%%%%%%%%%%%%%%%%

%%%%%%%%%%%%%%%%%%%%%%%%%%%%%%%%%%%%%%%%%%%%%%%%%%%%%%%%%%%%%%%%%%%%%%%%%%%%%%%
% Tables
%%%%%%%%%%%%%%%%%%%%%%%%%%%%%%%%%%%%%%%%%%%%%%%%%%%%%%%%%%%%%%%%%%%%%%%%%%%%%%%
\newpage

%------------  Form Factors in ISGW -------------------------------------------
\begin{table}
\caption{Form factors at $q^2=m_\pi^2\sim t_m$ $(q^2=m_\pi^2)$ in the ISGW (ISGW2) model.}
\begin{tabular}{c|ccc}
& $k$ & $b_+~({\rm GeV}^{-2})$ & $\F$ \\ \hline
$B\to a_2$&$0.058\sim0.320~~ (0.181)$&$-0.0034\sim-0.020~~ (-0.0040)$&$-0.031\sim-0.203~~ (0.078)$\\
$B\to f_2$&$0.055\sim0.320~~ (0.178)$&$-0.0032\sim-0.020~~ (-0.0039)$&$-0.030\sim-0.205~~ (0.076)$\\
$B\to f_2'$&$0.069\sim 0.320~~(0.212)$&$-0.0041\sim-0.020~~(-0.0042)$&$-0.035\sim-0.191~~ (0.103)$\\
$B\to K_2^*$&$0.064\sim 0.406~~(0.217)$&$-0.0037\sim-0.020~~(-0.0045)$&$-0.033\sim-0.111~~(0.102)$
\label{FISGW}
\end{tabular}
\end{table}

%%%%%%%%%%%%%%%%%%%%%%%%%%%%%%%%%%%%%%%%%%%%%%%%%%%%%%%%%%%%%%%%
%%%                       B to PT                            %%%
%%%%%%%%%%%%%%%%%%%%%%%%%%%%%%%%%%%%%%%%%%%%%%%%%%%%%%%%%%%%%%%%
\begin{table}
\caption{The branching ratios for $B \rightarrow PT$ decay modes
with $\Delta S =0$. The second, the third, and the fourth column
correspond to the cases for $\xi =0.1$, $\xi=0.3$, and $\xi=0.5$,
respectively.  $m_s (m_b) = 100$ MeV and $\gamma =65^0$ have been used.}
\smallskip
\begin{tabular}{c|ccc}
Decay mode   & ${\cal B}(10^{-7})~[\xi=0.1]$ & ${\cal
B}(10^{-7})~[\xi=0.3]$ & ${\cal B}(10^{-7})~[\xi=0.5]$
\\ \hline
%%%%%%%%%%%%%%%%%%%%%%%%%%%%%%%%%  charged modes
   $B^+ \rightarrow \pi^+ a_2^0$            & 29.73  & 26.02    & 22.56
\\ $B^+ \rightarrow \pi^+ f_2$              & 32.84  & 28.74    & 24.91
\\ $B^+ \rightarrow \pi^+ f_2^{\prime}$     & 0.42   & 0.37     & 0.32
\\ $B^+ \rightarrow \pi^0 a_2^+$            & 1.43   & 0.01     & 1.11
\\ $B^+ \rightarrow \eta a_2^+$             & 4.43   & 2.94     & 3.13
\\ $B^+ \rightarrow \eta^{\prime} a_2^+$    & 16.64  & 13.10    & 10.54
\\ $B^+ \rightarrow \bar K^0 K_2^{* +}$     & 0.0002  & 0.0004    & 0.002
\\
%%%%%%%%%%%%%%%%%%%%%%%%%%%%%%%%%  neutral modes
   $B^0 \rightarrow \pi^+ a_2^-$            & 55.79  & 48.82    & 42.32
\\ $B^0 \rightarrow \pi^0 a_2^0$            & 0.67   & 0.003    & 0.52
\\ $B^0 \rightarrow \pi^0 f_2$              & 0.74   & 0.003    & 0.58
\\ $B^0 \rightarrow \pi^0 f_2^{\prime}$     & 0.009  & 0.00004  & 0.007
\\ $B^0 \rightarrow \eta a_2^0$             & 2.08   & 1.38     & 1.47
\\ $B^0 \rightarrow \eta f_2$               & 2.30   & 1.52     & 1.62
\\ $B^0 \rightarrow \eta f_2^{\prime}$      & 0.03   & 0.02     & 0.02
\\ $B^0 \rightarrow \eta^{\prime} a_2^0$    & 7.81   & 6.15     & 4.95
\\ $B^0 \rightarrow \eta^{\prime} f_2$      & 8.63   & 6.80     & 5.47
\\ $B^0 \rightarrow \eta^{\prime} f_2^{\prime}$ & 0.11    & 0.09    & 0.07
\\$B^0 \rightarrow \bar K^0 K_2^{* 0}$          & 0.0002  & 0.0003  & 0.002
\label{PS0B}
\end{tabular}
\end{table}

%%%%%%%%%%%%%%%%%%%%%%%%%%%%%%%%%%%%%%%%%%%%%%%%%%%%%%%%%%%%%

\begin{table}
\caption{The CP asymmetries for $B \rightarrow PT$ decay modes
with $\Delta S =0$.  The definitions for the columns are the same
as those in Table II. }
\smallskip
\begin{tabular}{c|ccc}
Decay mode  & ${\cal A_{CP}}~[\xi=0.1]$ & ${\cal
A_{CP}}~[\xi=0.3]$ & ${\cal A_{CP}}~[\xi=0.5]$
\\ \hline
%%%%%%%%%%%%%%%%%%%%%%%%%%%%%%%%%  charged modes
   $B^+ \rightarrow \pi^+ a_2^0$                & $-0.01$  & $-0.01$   & $-0.01$
\\ $B^+ \rightarrow \pi^+ f_2$                  & $-0.01$  & $-0.01$   & $-0.01$
\\ $B^+ \rightarrow \pi^+ f_2^{\prime}$         & $-0.01$  & $-0.01$   & $-0.01$
\\ $B^+ \rightarrow \pi^0 a_2^+$                & $-0.04$  & $-0.55$   & 0.04
\\ $B^+ \rightarrow \eta a_2^+$                 & $-0.19$  & $-0.02$   & 0.18
\\ $B^+ \rightarrow \eta^{\prime} a_2^+$        & $-0.07$  & $-0.004$  & 0.07
\\ $B^+ \rightarrow \bar K^0 K_2^{* +}$         & 0        & 0         & 0
\\
%%%%%%%%%%%%%%%%%%%%%%%%%%%%%%%%%  neutral modes
   $B^0 \rightarrow \pi^+ a_2^-$                & $-0.01$  & $-0.01$   & $-0.01$
\\ $B^0 \rightarrow \pi^0 a_2^0$                & $-0.04$  & $-0.55$   & 0.04
\\ $B^0 \rightarrow \pi^0 f_2$                  & $-0.04$  & $-0.55$   & 0.04
\\ $B^0 \rightarrow \pi^0 f_2^{\prime}$         & $-0.04$  & $-0.55$   & 0.04
\\ $B^0 \rightarrow \eta a_2^0$                 & $-0.19$  & $-0.02$   & 0.18
\\ $B^0 \rightarrow \eta f_2$                   & $-0.19$  & $-0.02$   & 0.18
\\ $B^0 \rightarrow \eta f_2^{\prime}$          & $-0.19$  & $-0.02$   & 0.18
\\ $B^0 \rightarrow \eta^{\prime} a_2^0$        & $-0.07$  & $-0.004$  & 0.07
\\ $B^0 \rightarrow \eta^{\prime} f_2$          & $-0.07$  & $-0.004$  & 0.07
\\ $B^0 \rightarrow \eta^{\prime} f_2^{\prime}$ & $-0.07$  & $-0.004$  & 0.07
\\$B^0 \rightarrow \bar K^0 K_2^{* 0}$          & 0        & 0         & 0
\label{PS0A}
\end{tabular}
\end{table}

%%%%%%%%%%%%%%%%%%%%%%%%%%%%%%%%%%%%%%%%%%%%%%%%%%%%%%%%%%%%%%
\newpage
\begin{table}
\caption{The branching ratios  for $B \rightarrow PT$ decay modes
with $|\Delta S| =1$.  The definitions for the columns are the
same as those in Table II.}
\smallskip
\begin{tabular}{c|ccc}
Decay mode  & ${\cal B}(10^{-7})~[\xi=0.1]$ & ${\cal
B}(10^{-7})~[\xi=0.3]$ & ${\cal B}(10^{-7})~[\xi=0.5]$
\\ \hline
%%%%%%%%%%%%%%%%%%%%%%%%%%%%%%%%%  charged modes
   $B^+ \rightarrow K^+ a_2^0$                 & 3.56  & 3.11    & 2.70
\\ $B^+ \rightarrow K^+ f_2$                   & 3.94  & 3.44    & 2.98
\\ $B^+ \rightarrow K^+ f_2^{\prime}$          & 0.05  & 0.04    & 0.04
\\ $B^+ \rightarrow K^0 a_2^+$                 & 0.01  & 0.11    & 0.45
\\ $B^+ \rightarrow \pi^0 K_2^{* +}$           & 1.37  & 0.90    & 0.63
\\ $B^+ \rightarrow \eta K_2^{* +}$            & 2.56  & 0.31    & 0.28
\\ $B^+ \rightarrow \eta^{\prime} K_2^{* +}$   & 36.82 & 14.05   & 2.15
\\
%%%%%%%%%%%%%%%%%%%%%%%%%%%%%%%%%  neutral modes
   $B^0 \rightarrow K^+ a_2^-$                 & 6.68  & 5.84    & 5.06
\\ $B^0 \rightarrow K^0 a_2^0$                 & 0.003 & 0.05    & 0.21
\\ $B^0 \rightarrow K^0 f_2$                   & 0.003 & 0.05    & 0.23
\\ $B^0 \rightarrow K^0 f_2^{\prime}$          & 0.00004 & 0.0007  & 0.003
\\ $B^0 \rightarrow \pi^0 K_2^{* 0}$           & 1.27  & 0.84    & 0.58
\\ $B^0 \rightarrow \eta K_2^{* 0}$            & 2.37  & 0.29    & 0.26
\\ $B^0 \rightarrow \eta^{\prime} K_2^{* 0}$   & 34.15 & 13.04   & 1.99
\label{PS1B}
\end{tabular}
\end{table}

%%%%%%%%%%%%%%%%%%%%%%%%%%%%%%%%%%%%%%%%%%%%%%%%%%%%%%%%%%%%%%
\begin{table}
\caption{The CP asymmetries for $B \rightarrow PT$ decay modes
with $|\Delta S| =1$.  The definitions for the columns are the
same as those in Table II.}
\smallskip
\begin{tabular}{c|ccc}
Decay mode & ${\cal A_{CP}}~[\xi=0.1]$ & ${\cal A_{CP}}~[\xi=0.3]$
& ${\cal A_{CP}}~[\xi=0.5]$
\\ \hline
%%%%%%%%%%%%%%%%%%%%%%%%%%%%%%%%%  charged modes
   $B^+ \rightarrow K^+ a_2^0$              & 0.03  & 0.03  & 0.03
\\ $B^+ \rightarrow K^+ f_2$                & 0.03  & 0.03  & 0.03
\\ $B^+ \rightarrow K^+ f_2^{\prime}$       & 0.03  & 0.03  & 0.03
\\ $B^+ \rightarrow K^0 a_2^+$              & 0     & 0     & 0
\\ $B^+ \rightarrow \pi^0 K_2^{* +}$        & $-0.02$  & $-0.002$  & 0.04
\\ $B^+ \rightarrow \eta K_2^{* +}$         & 0.09   & 0.02   & 0.05
\\ $B^+ \rightarrow \eta^{\prime} K_2^{* +}$ & 0.01  & 0.001  & $-0.003$
\\
%%%%%%%%%%%%%%%%%%%%%%%%%%%%%%%%%  neutral modes
   $B^0 \rightarrow K^+ a_2^-$              & 0.03  & 0.03  & 0.03
\\ $B^0 \rightarrow K^0 a_2^0$              & 0     & 0     & 0
\\ $B^0 \rightarrow K^0 f_2$                & 0     & 0     & 0
\\ $B^0 \rightarrow K^0 f_2^{\prime}$       & 0     & 0     & 0
\\ $B^0 \rightarrow \pi^0 K_2^{* 0}$        & $-0.02$  & $-0.002$  & 0.04
\\ $B^0 \rightarrow \eta K_2^{* 0}$         & 0.09   & 0.02   & 0.05
\\ $B^0 \rightarrow \eta^{\prime} K_2^{* 0}$ & 0.01  & 0.001  & $-0.003$
\label{PS1A}
\end{tabular}
\end{table}

%%%%%%%%%%%%%%%%%%%%%%%%%%%%%%%%%%%%%%%%%%%%%%%%%%%%%%%%%%%%%%%%
%%%                       B to VT                            %%%
%%%%%%%%%%%%%%%%%%%%%%%%%%%%%%%%%%%%%%%%%%%%%%%%%%%%%%%%%%%%%%%%

\begin{table}
\caption{The branching ratios for $B \rightarrow VT$ decay modes
with $\Delta S =0$.
The definitions for the columns are the same as those in Table II.}
\smallskip
\begin{tabular}{c|ccc}
Decay mode   & ${\cal B}(10^{-7})~[\xi=0.1]$ & ${\cal
B}(10^{-7})~[\xi=0.3]$ & ${\cal B}(10^{-7})~[\xi=0.5]$
\\ \hline
%%%%%%%%%%%%%%%%%%%%%%%%%%%%%%%%%  charged modes
  $B^+ \rightarrow \rho^+ a_2^0$  & 83.90  & 73.42  & 63.65
\\  $B^+ \rightarrow \rho^+ f_2$  & 92.11  & 80.61  & 69.88
\\ $B^+ \rightarrow \rho^+ f_2^{\prime}$ & 1.18  & 1.03  & 0.89
\\  $B^+ \rightarrow \rho^0 a_2^+$ & 4.10  & 0.07  & 3.19
\\ $B^+ \rightarrow \omega a_2^+$  & 4.04  & 0.10  & 3.47
\\  $B^+ \rightarrow \phi a_2^+$   & 0.23  & 0.04  & 0.01
\\  $B^+ \rightarrow \bar K^{* 0} K_2^{* +}$ & 0.18 & 0.14 & 0.10
%%%%%%%%%%%%%%%%%%%%%%%%%%%%%%%%%%  neutral modes
\\  $ B^0 \rightarrow \rho^+ a_2^ -$ & 167.81 & 146.86 & 127.31
\\ $ B^0 \rightarrow \rho^0 a_2^0$ & 1.92  & 0.03  & 1.50
\\ $  B^0 \rightarrow \rho^0 f_2$  & 2.11  & 0.04  & 1.65
\\ $  B^0 \rightarrow \rho^0 f_2^{\prime}$ & 0.03  & 0.0005 & 0.02
\\ $ B^0 \rightarrow \omega a_2^0$ & 1.90  & 0.05  & 1.63
\\ $  B^0 \rightarrow \omega f_2$  & 2.08  & 0.05  & 1.79
\\ $ B^0 \rightarrow \omega f_2^{\prime}$  & 0.03 & 0.0006  & 0.02
\\ $ B^0 \rightarrow \phi a_2^0$   & 0.11 & 0.02 & 0.006
\\ $ B^0 \rightarrow \phi f_2$     & 0.12 & 0.02 & 0.006
\\ $ B^0 \rightarrow \phi f_2^{\prime}$ & 0.002 & 0.0002 & 0.0001
\\ $  B^0 \rightarrow \bar K^{* 0} K_2^{* 0}$ & 0.34 & 0.26 & 0.19
\label{VS0B}
\end{tabular}
\end{table}

%%%%%%%%%%%%%%%%%%%%%%%%%%%%%%%%%%%%%%%%%%%%%%%%%%%%%%%%%%%%%%%%%%%%%%%%

\begin{table}
\caption{The CP asymmetries for $B \rightarrow VT$ decay modes
with $\Delta S =0$.  The definitions for the columns are the same
as those in Table II. }
\smallskip
\begin{tabular}{c|ccc}
Decay mode  & ${\cal A_{CP}}~[\xi=0.1]$ & ${\cal
A_{CP}}~[\xi=0.3]$ & ${\cal A_{CP}}~[\xi=0.5]$
\\ \hline
%%%%%%%%%%%%%%%%%%%%%%%%%%%%%%%%%%%%%  charged modes
  $B^+ \rightarrow \rho^+ a_2^0$ & 0.02 & 0.02 & 0.02
\\  $B^+ \rightarrow \rho^+ f_2$ & 0.02 & 0.02 & 0.02
\\ $B^+ \rightarrow \rho^+ f_2^{\prime}$ & 0.02 & 0.02 & 0.02
\\  $B^+ \rightarrow \rho^0 a_2^+$ & 0.10 & 0.30 & $-0.09$
\\ $B^+ \rightarrow \omega a_2^+$  & $-0.01$   & $-0.21$   & 0.17
\\  $B^+ \rightarrow \phi a_2^+$   & 0 & 0 & 0
\\  $B^+ \rightarrow \bar K^{* 0} K_2^{* +}$  & 0 & 0 & 0
%%%%%%%%%%%%%%%%%%%%%%%%%%%%%%%%%%  neutral modes
\\  $ B^0 \rightarrow \rho^+ a_2^ -$ & 0.02 & 0.02 & 0.02
\\ $ B^0 \rightarrow \rho^0 a_2^0$  & 0.10 & 0.30 & $-0.09$
\\ $  B^0 \rightarrow \rho^0 f_2$   & 0.10 & 0.30 & $-0.09$
\\ $  B^0 \rightarrow \rho^0 f_2^{\prime}$ & 0.10 & 0.30 & $-0.09$
\\ $ B^0 \rightarrow \omega a_2^0$ & $-0.01$   & $-0.21$   & 0.17
\\ $  B^0 \rightarrow \omega f_2$  & $-0.01$   & $-0.21$   & 0.17
\\ $ B^0 \rightarrow \omega f_2^{\prime}$ & $-0.01$   & $-0.21$   & 0.17
\\ $ B^0 \rightarrow \phi a_2^0$  & 0 & 0 & 0
\\ $ B^0 \rightarrow \phi f_2$  & 0 & 0 & 0
\\ $ B^0 \rightarrow \phi f_2^{\prime}$  & 0 & 0 & 0
\\ $  B^0 \rightarrow \bar K^{* 0} K_2^{* 0}$   & 0 & 0 & 0
\label{VS0A}
\end{tabular}
\end{table}

%%%%%%%%%%%%%%%%%%%%%%%%%%%%%%%%%%%%%%%%%%%%%%%%%%%%%%%%%%%%%%
%%%%%%%%%%%%%%%%%%%%%%%%%%%%%%%%%%%%%%%%%%%%%%%%%%%%%%%%%%%%%%
\newpage
\begin{table}
\caption{The branching ratios  for $B \rightarrow VT$ decay modes
with $|\Delta S| =1$.  The definitions for the columns are the
same as those in Table II.}
\smallskip
\begin{tabular}{c|ccc}
Decay mode  & ${\cal B}(10^{-7})~[\xi=0.1]$ & ${\cal
B}(10^{-7})~[\xi=0.3]$ & ${\cal B}(10^{-7})~[\xi=0.5]$
\\ \hline
%%%%%%%%%%%%%%%%%%%%%%%%%%%%%%%%%%  charged modes
  $B^+ \rightarrow K^{* +} a_2^0$ & 20.82 & 18.52 & 16.35
\\  $B^+ \rightarrow K^{* +} f_2$ & 22.85 & 20.32 & 17.94
\\  $B^+ \rightarrow K^{* +} f_2^{\prime}$ & 0.28  & 0.25 & 0.22
\\  $B^+ \rightarrow   K^{* 0} a_2^+$ & 58.14 & 44.95 & 33.48
\\   $B^+ \rightarrow \rho^0 K_2^{* +}$ & 3.86  & 2.53  & 1.76
\\  $B^+ \rightarrow \omega K_2^{* +}$  & 23.92 & 1.12  & 7.89
\\  $B^+ \rightarrow \phi K_2^{* +}$    & 5.57  & 21.80 & 48.81
%%%%%%%%%%%%%%%%%%%%%%%%%%%%%%%%%%%  neutral modes
\\ $ B^0 \rightarrow K^{* +} a_2^-$  & 39.09 & 34.77 & 30.71
\\ $ B^0 \rightarrow  K^{* 0} a_2^0$ & 27.28 & 21.09 & 15.71
\\ $ B^0 \rightarrow  K^{* 0} f_2$   & 29.93 & 23.14 & 17.23
\\  $ B^0 \rightarrow  K^{* 0} f_2^{\prime}$ & 0.37  & 0.29  & 0.21
\\ $ B^0 \rightarrow \rho^0  K_2^{* 0}$  & 3.58  & 2.35  & 1.63
\\  $ B^0 \rightarrow \omega  K_2^{* 0}$ & 22.21 & 1.04  & 7.32
\\   $ B^0 \rightarrow \phi  K_2^{* 0}$  & 5.17  & 20.24 & 45.32
\label{VS1B}
\end{tabular}
\end{table}

%%%%%%%%%%%%%%%%%%%%%%%%%%%%%%%%%%%%%%%%%%%%%%%%%%%%%%%%%%%%%%
%%%%%%%%%%%%%%%%%%%%%%%%%%%%%%%%%%%%%%%%%%%%%%%%%%%%%%%%%%%%%%
\begin{table}
\caption{The CP asymmetries for $B \rightarrow VT$ decay modes
with $|\Delta S| =1$.  The definitions for the columns are the
same as those in Table II.}
\smallskip
\begin{tabular}{c|ccc}
Decay mode & ${\cal A_{CP}}~[\xi=0.1]$ & ${\cal A_{CP}}~[\xi=0.3]$
& ${\cal A_{CP}}~[\xi=0.5]$
\\ \hline
%%%%%%%%%%%%%%%%%%%%%%%%%%%%%%  charged modes
  $B^+ \rightarrow K^{* +} a_2^0$ & $-0.33$ & $-0.32$ & $-0.32$
\\ $B^+ \rightarrow K^{* +} f_2$  & $-0.33$ & $-0.32$ & $-0.32$
\\ $B^+ \rightarrow K^{* +} f_2^{\prime}$ & $-0.33$ & $-0.32$ & $-0.32$
\\ $B^+ \rightarrow   K^{* 0} a_2^+$  & 0 & 0 & 0
\\ $B^+ \rightarrow \rho^0 K_2^{* +}$ & 0.02 & 0.002 & $-0.04$
\\  $B^+ \rightarrow \omega K_2^{* +}$ & $-0.05$ & $-0.007$ & $-0.10$
\\  $B^+ \rightarrow \phi K_2^{* +}$  & 0 & 0 & 0
%%%%%%%%%%%%%%%%%%%%%%%%%%%%%%  neutral modes
\\ $ B^0 \rightarrow K^{* +} a_2^-$ & $-0.33$ & $-0.32$ & $-0.32$
\\ $ B^0 \rightarrow  K^{* 0} a_2^0$  & 0 & 0 & 0
\\ $ B^0 \rightarrow  K^{* 0} f_2$    & 0 & 0 & 0
\\  $ B^0 \rightarrow  K^{* 0} f_2^{\prime}$ & 0 & 0 & 0
\\ $ B^0 \rightarrow \rho^0  K_2^{* 0}$ & 0.02 & 0.002 & $-0.04$
\\  $ B^0 \rightarrow \omega  K_2^{* 0}$ & $-0.05$ & $-0.007$ & $-0.10$
\\   $ B^0 \rightarrow \phi  K_2^{* 0}$ & 0 & 0 & 0
\label{VS1A}
\end{tabular}
\end{table}

%%%%%%%%%%%%%%%%%%%%%%%%%%%%%%%%%%%%%%%%%%%%%%%%%%%%%%%%%%%%%%%
%%%%%%%%%%%%%%%%%%%%%%%%%%%%%%%%%%%%%%%%%%%%%%%%%%%%%%%%%%%%%%%
\begin{table}
\caption{Ratios of the branching ratios for $B \to VT$ and for $B
\to PT$ decay modes, where $V$ and $P$ have identical quark
content. The second column corresponds to the case for $m_s(m_b) = 100$
MeV and $\gamma =65^0$.  The values of $\xi$ vary from 0.1 to
0.5~. }
\smallskip
\begin{tabular}{cc}
Ratio & $m_s = 100$ MeV, $\gamma =65^0$
\\ \hline
%%%%%%%%%%%%%%%%%%%%%%%%%%%%%%  Delta S = 0
${\cal B}(B^+ \rightarrow \rho^+ a_2^0$) / ${\cal B}(B^+
\rightarrow \pi^+ a_2^0$)  & 2.82
\\  ${\cal B}(B^+ \rightarrow \rho^+ f_2$) / ${\cal B}(B^+ \rightarrow
\pi^+ f_2$)  & 2.80
\\ ${\cal B}(B^0 \rightarrow \rho^+ a_2^-$) / ${\cal B}(B^0 \rightarrow
\pi^+ a_2^-$)  & 3.01
%%%%%%%%%%%%%%%%%%%%%%%%%%%%%%  |Delta S| = 1
\\ ${\cal B}(B^+ \rightarrow K^{* +} a_2^0$) / ${\cal B}(B^+ \rightarrow
K^+ a_2^0$)  & 5.85$-$6.06
\\ ${\cal B}(B^+ \rightarrow K^{* +} f_2$) / ${\cal B}(B^+ \rightarrow
K^+ f_2$)  & 5.80$-$6.02
\\ ${\cal B}(B^0 \rightarrow K^{* +} a_2^-$) / ${\cal B}(B^0 \rightarrow
K^+ a_2^-$)  & 5.85$-$6.07
\label{RVP}
\end{tabular}
\end{table}

%%%%%%%%%%%%%%%%%%%%%%%%%%%%%%%%%%%%%%%%%%%%%%%%%%%%%%%%%%%%%%%
%%%%%%%%%%%%%%%%%%%%%%%%%%%%%%%%%%%%%%%%%%%%%%%%%%%%%%%%%%%%%%%

\end{document}